\documentclass[preprint,nofootinbib,superscriptaddress]{revtex4-1}

\usepackage{amssymb, amsmath,framed}
\usepackage{cancel}
\usepackage{amsfonts}
\usepackage[T1]{fontenc}
\usepackage{bm}
\usepackage{color}
\usepackage{graphicx}
\DeclareSymbolFont{operators}{OT1}{cmr}{m}{n}
\DeclareSymbolFont{letters}{OML}{cmm}{m}{it}
\DeclareSymbolFont{symbols}{OMS}{cmsy}{m}{n}
\DeclareSymbolFont{largesymbols}{OMX}{cmex}{m}{n}

\usepackage{hyperref}
\hypersetup{
    bookmarks=true,         
    unicode=false,          
    pdftoolbar=true,        
    pdfmenubar=true,        
    pdffitwindow=false,     
    pdfstartview={FitH},    
    pdfsubject={Plasma fluid theory},   
    pdfproducer={GNU Make}, 
    pdfkeywords={keyword1} {key2} {key3}, 
    pdfnewwindow=true,      
    colorlinks=true,        
    linkcolor=blue,      
    citecolor=blue,         
    filecolor=blue,      
    urlcolor=blue           
}
\usepackage{array}

\makeatother

\begin{document}
\title{Physical constraints on the likelihood of life on exoplanets}

\author{Manasvi Lingam}
\email{manasvi@seas.harvard.edu}
\affiliation{John A. Paulson School of Engineering and Applied Sciences, Harvard University, 29 Oxford St, Cambridge, MA 02138, USA}
\affiliation{Harvard-Smithsonian Center for Astrophysics, 60 Garden St, Cambridge, MA 02138, USA}
\author{Abraham Loeb}
\email{aloeb@cfa.harvard.edu}
\affiliation{Harvard-Smithsonian Center for Astrophysics, 60 Garden St, Cambridge, MA 02138, USA}

\date{}

\begin{abstract}
One of the most fundamental questions in exoplanetology is to determine whether a given planet is habitable. We estimate the relative likelihood of a planet's propensity towards habitability by considering key physical characteristics such as the role of temperature on ecological and evolutionary processes, and atmospheric losses via hydrodynamic escape and stellar wind erosion. From our analysis, we demonstrate that Earth-sized exoplanets in the habitable zone around M-dwarfs seemingly display much lower prospects of being habitable relative to Earth, owing to the higher incident ultraviolet fluxes and closer distances to the host star. We illustrate our results by specifically computing the likelihood (of supporting life) for the recently discovered exoplanets, Proxima b and TRAPPIST-1e, which we find to be several orders of magnitude smaller than that of Earth.
\end{abstract}

\maketitle

\section{Introduction} \label{SecIntro}
The field of exoplanetary science has witnessed many rapid and exciting advances over the past couple of decades. With the number of discovered exoplanets now numbering in the thousands \citep{WF15}, there has been a great deal of interest in identifying planets that are habitable, i.e. potentially capable of bearing life \citep{KaCa03,CH05,Lam09,Cock16}. To this end, most of the current discoveries have been centered around planets orbiting M-dwarfs since they are more abundant, and easier to detect \citep{SKS07,Tart07,SBJ16}. It is now believed that there are $\sim 10^{10}$ habitable planets in our own Galaxy \citep{KKRH14,CD15}.

In recent times, several major discoveries have provided additional impetus to studies of habitability. The first was the discovery of an Earth-mass exoplanet in the habitable zone (HZ) - the region capable of sustaining liquid water on the planet's surface - around Proxima Centauri, the nearest star to our Solar system at only $4.2$ light years away \citep{AE16}. There are plans already underway to explore this planet, dubbed Proxima b, via a flyby mission.\footnote{\url{https://breakthroughinitiatives.org/Initiative/3}} The second major advance was the discovery of at least seven Earth-sized planets transiting the ultracool dwarf star TRAPPIST-1 at a distance of $39.5$ light years \citep{Gill16,Gill17}. The presence of three planets in the HZ around TRAPPIST-1 therefore presents a unique opportunity for studying multiple planets that may host life \citep{LL17}. The discovery of the temperate super-Earth LHS 1140b transiting an M-dwarf at a distance of $39$ light years also merits a mention \citep{DID17}.

It is worth emphasizing that the existence of a planet in the HZ does \emph{not} imply life exists on the planet, or even that it will necessarily be able to support a biosphere. Most of the current assessments of habitable planets have tended to focus on superficial metrics which can be misleading, as rightly pointed out in \citet{SG16} and \citet{Task17}. These metrics evaluate the degree of similarity between certain, physically relevant, parameters of a given exoplanet and the Earth, and have led to unfortunate misconceptions that planets with higher values (of the similarity indices) are automatically more habitable.

In this paper, we attempt to advance the assessment of life-bearing planets by identifying physical processes that play a major role in governing habitability through the formulation of likelihood functions that depend on basic planetary and stellar parameters. By doing so, we expect to pave the way towards understanding the complex relationship between the aforementioned parameters and the likelihood of a planet hosting life in reality \citep{LC12,Cock16}.

\section{The role of planetary temperature} \label{SecTemp}
We begin by outlining the centrality of temperature in regulating a diverse array of ecological and evolutionary parameters and processes \citep{CB87,BGASW,KinHu08,DTH08,Ang09,DWH10} by adopting the general premise that biochemical reactions analogous to metabolism are universal on life-bearing exoplanets \citep{Pace01,BRC04,Ball08}. Subsequently, we construct a global likelihood function for these processes that depends on the planetary temperature.

\subsection{The Metabolic Theory of Ecology and Temperature}
We adopt the tenets of the Metabolic Theory of Ecology (MTE) that relies on the assumption that the metabolic rate $B$ of organisms plays a major role in governing macroecological processes. The reader may consult \citet{WBE01,BGASW,AW04,CF04,Marq05,WeBo05,Sav08,IC10,Price12,KW12,HuCa14,Marq14} for comprehensive reviews, assessments and critiques of this field. We have chosen to work with this model since it attempts to quantify important ecological patterns and parameters by adopting a mechanistic perspective based on generic physical and chemical considerations, without the necessity for invoking complex (and specific) biological factors. Naturally, this approach has attracted a fair amount of criticism as summarized in the aforementioned references. Nevertheless, we will operate under the premise that the basic underpinnings of the MTE are valid for ecosystems on other planets, at least for those capable of sustaining life-as-we-know-it.

The MTE is founded on the principle that the metabolic rate, which serves as its cornerstone, scales as,
\begin{equation} \label{KLTemp}
    B \propto m^{3/4}\, \exp\left(-\frac{E}{k_B T}\right),
\end{equation}
where $m$ is the mass of the organism, $k_B$ is the Boltzmann constant, $T$ is the absolute temperature and $E$ is the average activation energy which is determined by considering the appropriate rate-limiting step in metabolism \citep{GBWSC}. Now, suppose that one wished to formulate a ``mean'' metabolic rate $\bar{B}$ across all species. This can be done by introducing the distribution function for the number of individuals with a given mass $N(m)$ as follows,
\begin{equation} \label{AvMBR}
    \bar{B} = \frac{\int B(m,T) N(m)\,dm}{\int N(m)\,dm} \propto \exp\left(-\frac{E}{k_B T}\right).
\end{equation}
The last scaling follows if $E(m,T) \approx E(T)$, i.e. provided that the activation energy displays a weak dependence on the organism's mass.\footnote{We have also assumed that the temperature dependence of the distribution function $N$ is minimal compared to its mass dependence.} This appears to be a fairly robust assumption on Earth, since $E$ falls within a fairly narrow band of energies ranging between $0.6$-$0.7$ eV, for unicellular organisms, plants, ectotherms and endotherms \citep{GBWSC,BGASW,DPS11}. It was pointed out in \citet{Gillo06} that the mean value of $E=0.65$ eV is nearly equal to the average activation energy of $0.66$ eV that arises from the rate of ATP synthesis in isolated mitochondria; this value of $E$ has also been explored in \citet{YCC12}.

One of the central predictions of the MTE is that several ecological parameters are regulated by the metabolic rate, and are thus expected to depend on the temperature via the Boltzmann factor inherent in Eq. (\ref{KLTemp}). Examples of these parameters, which have been studied empirically, include:
\begin{itemize}
\item The production and turnover of biomass, the rate of biological energy flux per unit area, and the metabolic balance of ecosystems \citep{TVLH93,Enq03,AGB05,LUS06,AlGil09,YD10}.
\item The maximal rates of population growth and molecular evolution \citep{SavGi04,BGASW,Gill05}, and the reciprocal developmental time \citep{Gill02,OC07,EK07}.
\item The rates of genetic divergence and speciation, species diversity and coexistence, and trophic interactions \citep{ABG02,AlGil06,DPS14}. 
\item Higher biodiversity is predicted to be prevalent in habitats with hotter average temperatures \citep{Marq05,Fuhr08,WBTF09,Bro14}. This pattern has been proven to be valid on Earth, which is characterized by distinctive latitudinal gradients in species richness \citep{Ros95,Gas00}, although other factors, such as the degree of precipitation, also play an important, but possibly sub-dominant, role \citep{Moles14}.
\end{itemize}
The Boltzmann factor dependence implies that all of the above quantities are expected to monotonically increase with temperature. This is closely related to the notion that ``\emph{Hotter is better}'', which posits that a higher value of $T$ is correlated with enhanced growth, fitness and diversity \citep{Thomp42,KinHu08,May12}, up to a particular limit.

At this stage, some important caveats are in order. The first stems from critiques concerning the validity and interpretation of the MTE \citep{Clarke06,OCon07,DHT08,Glaz15}. In addition, the monotonically increasing trend with temperature cannot continue \emph{ad infinitum} since thermal adaptation breaks down beyond a certain point \citep{Ang09,DGS11,Sch15}. Moreover, there exist several important and subtle ambiguities in resolving the exact relationship between temperature and the aforementioned traits \citep{CF04,CR08,King09}. Lastly, our analysis has presupposed a steady-state temperature, but rapid fluctuations can engender mass extinctions and irreversible changes in the biosphere \citep{PuHe00,Barn12,FraSu14}.

\subsection{The temperature-dependent likelihood function}
Based on the preceding discussion, we introduce the likelihood factor for biodiversity which has been adapted from Eq. (\ref{AvMBR}),
\begin{equation} \label{SharpCutoff}
\exp\left(-\frac{E}{k_B T}\right)\, \theta\left(T-T_L\right)\theta\left(T_U - T\right),  
\end{equation}
where $E$ and $T$ are now taken to be the mean activation energy and temperature respectively. $\theta$ is the Heavyside function which ensures that the likelihood becomes zero for $T < T_L$ and $T > T_U$. 

The limits of Earth-based lifeforms fall within $262$ K and $395$ K \citep{McK14,RoMa01}, while the corresponding range for photosynthetic lifeforms is narrower since the upper bound is lowered to $348$ K \citep{KSGB}. In Eq. (\ref{SharpCutoff}), note that $T$ corresponds to the average surface temperature of the planet in our subsequent analysis, because we are interested in quantifying the likelihood on a \emph{planetary} scale. Thus, instead of $T$ representing the temperature of the local habitat (for ectotherms) or the average internal temperature (for endotherms), we have replaced it by the overall planetary surface temperature. Naturally, a limitation of this methodology is that it represents a coarse-grained estimate that smears out local effects.\footnote{Our approach is somewhat analogous to that of metapopulation ecology, wherein each ``unit'' is a population patch comprising of several individuals, while the metapopulation has been typically envisioned as a ``population of populations'' \citep{Lev69,Han98}.} 

We are now in a position to construct the ``normalized'' (Earth-referenced) temperature-based likelihood with respect to the Earth, \begin{equation} \label{TempLike}
\mathcal{P}_T = \exp\left[-26.7\left(\delta-1\right)\right]\, \theta\left(T-T_L\right)\theta\left(T_U - T\right),
\end{equation}
where $T_U$ and $T_L$ should be interpreted as the limits over which the Boltzmann factor is valid. Clearly, the simple ansatz exemplified by Eq. (\ref{SharpCutoff}) corresponds to sharp (discontinuous) cutoffs, and we refer the reader to \citet{CORMR12,DPS14,Sch15,COMBROR}, wherein more sophisticated and realistic variants have been outlined. The temperature range $T_L < T < T_U$ is not expected to exceed the Earth-based photosynthesis limits for life-as-we-know-it.\footnote{The assumption can be easily relaxed, and, indeed, a rich array of ecosystems based on alternative biochemistries have been extensively investigated \citep{Bai04,SI08}.} We have also introduced the auxiliary parameter $\delta$ in Eq. (\ref{TempLike}),
\begin{equation} \label{deldef}
\delta = \frac{E}{E_\oplus}\frac{T_\oplus}{T} \sim \frac{T_\oplus}{T},
\end{equation}
where $E_\oplus = 0.66$ eV \citep{DPS11} and $T_\oplus = 287$ K are the corresponding values for the Earth. The last relation follows from $E \sim E_\oplus$ since we only consider life-as-we-know-it in our present analysis. If we substitute $T = 218$ K for Mars, we find that $\mathcal{P}_T = 0$ because of the Heavyside function. In contrast, the equatorial temperatures on Mars can exceed $T_L$, thereby giving rise to a finite value of $\mathcal{P}_T$ \emph{locally}. If we consider Venus instead, it is evident that $\mathcal{P}_T = 0$ since $T \gg T_U$. 

Thus, to conclude, we have hypothesized that Eq. (\ref{TempLike}) quantifies the likelihood of complex life-sustaining processes as a function of the planet's surface temperature.\footnote{In Section 2.3 of \citet{Ho07}, the temperature dependence of the metabolic rate was briefly discussed from the perspective of bioenergetics.} This function serves as a proxy for the (relative) metabolic rate, which, in turn, has been hypothesized to regulate important biological parameters such as the species diversity, biological fluxes, and the rates of speciation and growth to name a few. Hence, if abiogenesis had been successfully initiated on a particular planet, the chances of complex life emerging are expected to be correspondingly greater for higher temperatures since: (i) the rates of evolution and speciation are enhanced, and (ii) a greater diversity of species are potentially sustainable. Hence, this metric arguably constitutes a more sophisticated variant of understanding the likelihood of macroecological processes on exoplanets. However, we must reiterate that this likelihood function is \emph{not} synonymous with a planet being habitable since there are myriad factors involved in the latter. Furthermore, Eq. (\ref{TempLike}) does not quantify the likelihood of abiogenesis, and the existence of the aforementioned ecological and evolutionary processes is obviously contingent upon life successfully originating (and diversifying) on the planet.

Although we have not directly estimated the prospects for abiogenesis on exoplanets, we wish to point out that several studies have presented empirical and theoretical evidence favoring a high-temperature origin of life \citep{Pace91,GTBB03,MBRR08,ANY13}, although many factors still remain poorly known \citep{MiLa95}. If life did indeed originate in a high-temperature environment, perhaps the likelihood of abiogenesis could exhibit a Boltzmann factor dependence on the temperature akin to Eq. (\ref{AvMBR}), thereby favoring thermophilic ancestral lifeforms \citep{Weiss16}. However, an important point worth highlighting is that the temperature alluded to when discussing abiogenesis always represents the \emph{in situ} value (for e.g., at hydrothermal vents), and not the global planetary temperature.

\section{The role of atmospheric escape}
Next, we explore the constraints on the likelihood (of a planet being habitable) that are set by atmospheric escape. Before doing so, a few general observations are in order.

The phase diagram of water requires external pressure in order for liquid water to emerge upon warming solid ice \citep{McK14}. It is therefore widely presumed that complex, surface-based, organic chemistry corresponding to life-as-we-know-it necessitates the existence of an atmosphere \citep{Lam09,Cock16}. However, atmospheres can be eroded through a diverse array of processes such as thermal escape, photochemical escape, and multiple non-thermal mechanisms such as sputtering \citep{John08,Seager10,Lammer13}.\footnote{In addition, planets could lose a significant fraction of their atmospheres due to X-ray and ultraviolet irradiation from supermassive black holes \citep{FoLo17}.} On the other hand, they can also be replenished through volcanism, giant impacts and evaporation of oceans \citep{KaCa03}. If the associated timescales for the escape processes are `fast', there may not exist sufficient time for complex life to emerge and evolve. Thus, we shall suppose henceforth that the timescales for atmospheric escape, which can be quantified in some instances, will serve as effective constraints for determining the inclination towards habitability of a given exoplanet relative to Earth.

In actuality, there are several important and distinct timescales that must be taken into account in conjunction with the planet's age $\left(t_P\right)$. Some of the notable ones are as follows:
\begin{itemize}
    \item The characteristic timescale(s) involved in the depletion of the planetary atmosphere $\left(t_\ell\right)$, which will be the focus of this paper.
    \item The minimum timescale required for abiogenesis - the origin of life (OOL) - to commence $\left(t_{OOL}\right)$, since the likelihood of life arising on the planet is zero for $t_P < t_{OOL}$. As per current evidence, it seems plausible that $t_{OOL} \lesssim 500$ Myr on Earth \citep{BBHM,Dod17} although, in light of the many uncertainties involved, this timescale ought not be perceived as being definitive.
    \item The timescale over which the planet will be subjected to the extended pre-main-sequence (pre-MS) phase $\left(t_{PMS}\right)$, since this can adversely impact habitability (extreme water loss) especially when dealing with exoplanets in the HZ of M-dwarfs \citep{RaKa14,LuBa15,TI15}.
    \item The duration of time taken for a given planet to ``enter'' the outer edge of the HZ $\left(t_{OHZ}\right)$, and to finally ``exit'' the inner edge of the HZ $\left(t_{IHZ}\right)$, since the HZ itself evolves over time \citep{RCOW}. By construction,  $t_P < t_{OHZ}$ or $t_P > t_{IHZ}$ imply the planet would not be habitable (in the conventional sense).
\end{itemize}
Thus, we shall implicitly restrict our attention to planets where $t_P > t_{OOL}$,  $t_{OHZ} < t_P < t_{IHZ}$, and the pre-MS phase has not rendered the planet uninhabitable. 

\subsection{Timescales for atmospheric escape}
As noted earlier, there are several mechanisms that lead to atmospheric losses. In our analysis, we shall consider two dominant causes, namely, hydrodynamic escape and stellar wind stripping. We do not evaluate the extent of Jeans escape, as it does not play a major role in facilitating escape of heavier molecules (such as O$_2$ and CO$_2$) from Earth-like planets in the HZ \citep{Lammer13}.

For hydrodynamic escape, we rely upon the assumption of energy-limited escape \citep{Seager10}, enabling us to estimate the lifetime of the planet's atmosphere as,
\begin{equation} \label{tHD}
    t_{HD} \sim \frac{G M_p}{\pi R_p^3} \frac{M_{atm}}{\beta \eta \langle{F_{EUV}}\rangle},
\end{equation}
where $M_{atm}$, $M_p$ and $R_p$ are the mass of the atmosphere, mass and radius of the planet respectively; $\langle{F_{EUV}}\rangle$ is the average extreme ultraviolet (EUV) flux  while $\beta$ and $\eta$ are phenomenological parameters. Further details concerning the derivation of Eq. (\ref{tHD}) can be found in Chapter 4 of \citet{Seager10} and Section 2 of \citet{OA16}. Other forms of UV-driven atmospheric loss include recombination-limited and photon-limited escape \citep{OA16}. 

Next, we consider stripping of the atmosphere by the stellar wind. The associated timescales are given by $t_{MSW}$ and $t_{UMSW}$ for the magnetized and unmagnetized cases (with and without an intrinsic dipole field). Note that we are interested in the \emph{fastest} timescale for atmospheric loss, thereby leading us to $t_\ell = \mathrm{min} \{t_{HD},t_{UMSW}\}$ or $t_\ell = \mathrm{min} \{t_{HD},t_{MSW}\}$ depending on the situation. However, the escape rates for magnetized planets are usually expected to be somewhat lower \citep{Lammer13,Ehl16,DLMC}, implying that $t_{UMSW} < t_{MSW}$. 

In fact, for most exoplanets in the HZ around M-dwarfs, the dynamic pressure exerted by the stellar wind is so great that the additional shielding offered by the planet's magnetic field $B_P$ is relatively unimportant \citep{GDC16,DLMC,Aira17}, provided that $B_P$ is not anomalously high \citep{Vid13}. By utilizing standard dynamo scaling laws \citep{Christ10}, it can be shown that $B_P$ will not be very large if the convected heat flux in the planetary core is not significantly higher than that of the Earth. In fact, tidally locked exoplanets, expected to be fairly common in the HZ around M-dwarfs, are typically associated with weak magnetic moments \citep{Khoda07,SBJ16}.

Based on the above set of arguments, we may henceforth adopt $t_{UMSW} \sim t_{MSW}$ for certain M-dwarf exoplanets. The timescale of atmospheric loss for unmagnetized planets $\left(t_{SW}\right)$ is given by,
\begin{equation} \label{tSW}
t_{SW} \sim \frac{2}{\alpha} \frac{M_{atm}}{\dot{M}_\star} \left(\frac{a}{R_P}\right)^2,
\end{equation}
where $a$ is the semi-major axis of the planet, $\dot{M}_\star$ is the stellar mass loss rate, and $\alpha$ is the entrainment efficiency that is treated as a constant \citep{ZSR10}. The above formula was verified to be fairly accurate by means of numerical simulations in \citet{DJL17}.

\subsection{Contribution of atmospheric loss to the likelihood function}
Following our preceding discussion, the atmospheric loss is given by $t_\ell = \mathrm{min}\{t_{HD},t_{SW}\}$, where $t_{HD}$ and $t_{SW}$ are given by Eqs. (\ref{tHD}) and (\ref{tSW}) respectively. We define the normalized (Earth-referenced) likelihood of this timescale in the following manner:
\begin{equation} \label{AtLF}
\mathcal{P}_A = \frac{t_\ell}{t_{\oplus}},
\end{equation}
where $t_{\oplus}$ is the corresponding value of $t_\ell$ upon substituting Earth's parameters. When $t_{HD} < t_{SW}$, we find that Eq. (\ref{AtLF}) simplifies to,
\begin{equation} \label{HDLike}
\mathcal{P}_A(HD) = \left(\frac{P_s}{1\,\mathrm{atm}}\right) \left(\frac{R_P}{R_\oplus}\right) \left(\frac{\langle{F_{EUV}}\rangle}{\langle{F_{\oplus}}\rangle}\right)^{-1},
\end{equation}
where $P_s = g M_{atm}/(4\pi R_P^2)$ is the surface pressure of the atmosphere, $F_{\oplus}$ is the value of $\langle{F_{EUV}}\rangle$ for the Earth, and we have employed the mass-radius relation $M/M_\oplus \sim \left(R/R_\oplus\right)^{3.7}$ \citep{VOCS06,ZSJ16}. Similarly, when $t_{HD} > t_{SW}$, it is easy to verify that Eq. (\ref{AtLF}) can be expressed as,
\begin{equation} \label{SWLike}
\mathcal{P}_A(SW) = \left(\frac{P_s}{1\,\mathrm{atm}}\right) \left(\frac{a}{1\,\mathrm{AU}}\right)^2 \left(\frac{R_P}{R_\oplus}\right)^{-1.7} \left(\frac{\dot{M}_\star}{\dot{M}_\odot}\right)^{-1},
\end{equation}
where $\dot{M}_\odot$ is the Sun's mass-loss rate. In general, the mass-loss rate displays a complex empirical dependence on stellar parameters, for instance the star's mass, activity, age, and rotation rate \citep{CS11}. We shall adopt the scaling relation proposed in \citet{JGBL15} for low-mass stars (although its validity remains somewhat uncertain for $M_\star < 0.4 M_\odot$ and $M_\star > 1.1 M_\odot$):
\begin{equation}
    \frac{\dot{M}_\star}{\dot{M}_\odot} = \left(\frac{R_\star}{R_\odot}\right)^2 \left(\frac{\Omega_\star}{\Omega_\odot}\right)^{1.33} \left(\frac{M_\star}{M_\odot}\right)^{-3.36},
\end{equation}
where $R_\star$, $\Omega_\star$ and $M_\star$ are the stellar radius, rotation rate and mass respectively. When we substitute the values for Proxima Centauri, we find that $\dot{M}_\star \approx 4.8 \dot{M}_\odot$, which is fairly close to $\dot{M}_\star \approx \dot{M}_\odot$ obtained from simulations \citep{GDC16}. The above expression is based on the assumption that the star does not rotate rapidly,\footnote{A rapid rotation rate corresponds to a value that is close to the saturation estimate of $\Omega_\mathrm{sat} = 15 \Omega_\odot \left(M_\star/M_\odot\right)^{2.3}$ \citep{JGBL15}.} since the mass-loss rate would otherwise attain a saturation value with the dependence $\dot{M}_\star \propto M_\star^{1.3}$ in that regime \citep{JGBL15}.

A few general observations can be drawn from Eqs. (\ref{HDLike}) and (\ref{SWLike}). The likelihood function is linearly proportional to the surface pressure (which itself is related to $M_{atm}$), implying that planets with relatively massive atmospheres are conducive to being superhabitable, in agreement with \citet{VMSP} and \citet{HeArm14}. Secondly, we observe that exoplanets in the HZ around M-dwarfs are generally subject to higher values of $\langle{F_{EUV}}\rangle$ compared to the Earth \citep{FFL13}, and are also located much closer \citep{Kop13}. When combined with Eqs. (\ref{HDLike}) and (\ref{SWLike}), these facts imply that $\mathcal{P}_A$ for such exoplanets is likely to be much lower than unity. 

We have seen that $t_\ell$ quantifies the timescale over which the atmosphere is present. Over this duration, it is worth highlighting that species diversity itself increases over time. The enhancement of biodiversity has been predicted to obey logistic growth \citep{PuHe00,Ben09}, which can, in some instances, be loosely visualized as a linear function during the growing phase prior to saturation. Thus, one could, perhaps, also envision Eq. (\ref{AtLF}) as a heuristic measure of the maximal species diversity relative to Earth. In turn, a planetary ecosystem with higher biodiversity would be typically associated with greater stability and multifunctionality \citep{Hoop05,Card12}, although the subtleties inherent in analyses of diversity-stability relationships should be duly recognized \citep{IC07}.

Lastly, we observe that $\mathcal{P}_A = 0$ when $t_\ell < t_{OOL}$ since the planet's atmosphere is lost prior to the onset of abiogenesis. Similarly, if $t_\ell > t_{HZ}:=t_{IHZ}-t_{OHZ}$, one must replace $t_\ell$ in Eq. (\ref{AtLF}) with $t_{HZ}$ since the latter would become the critical timescale in this regime.

\begin{table*}
\begin{minipage}{126mm}
\caption{The likelihood function relative to Earth for different HZ exoplanets}
\label{Tab1}
\begin{tabular}{|c|c|c|c|c|c|}
\hline 
Planet & $\mathcal{P}_T$ & $\mathcal{P}_A(HD)$ & $\mathcal{P}_A(SW)$ & $\mathcal{P}(HD)$ & $\mathcal{P}(SW)$\tabularnewline
\hline 
\hline 
Earth & $1$ & $1$ & $1$ & $1$ & $1$\tabularnewline
\hline 
Proxima b & $0.13$ & $3 \times 10^{-2}$ & $2 \times 10^{-3}$ & $3.9 \times 10^{-3}$ & $2.6 \times 10^{-4}$\tabularnewline
\hline 
TRAPPIST-1e & $0.93$ & $4.5 \times 10^{-2}$ & $9.1 \times 10^{-4}$ & $4.2 \times 10^{-2}$ & $8.5 \times 10^{-4}$\tabularnewline
\hline 
TRAPPIST-1f & $0$ & $8.7 \times 10^{-2}$ & $1.3 \times 10^{-3}$ & $0$ & $0$\tabularnewline
\hline 
TRAPPIST-1g & $0$ & $0.14$ & $1.7 \times 10^{-3}$ & $0$ & $0$\tabularnewline
\hline 
\end{tabular}
\medskip

{\bf Notes:} $\mathcal{P}_A(HD)$ and $\mathcal{P}_A(SW)$ are given by Eqs. (\ref{HDLike}) and (\ref{SWLike}) respectively, while $\mathcal{P}(HD) = \mathcal{P}_T \cdot \mathcal{P}_A(HD)$ and $\mathcal{P}(SW) = \mathcal{P}_T \cdot \mathcal{P}_A(SW)$. The stellar mass-loss rate for Proxima Centauri is from \citet{GDC16} (see also \citet{WoLi01}) and the corresponding value for TRAPPIST-1 has been assumed to be approximately equal to that of Proxima Centauri. The stellar and planetary parameters have been tabulated in \citet{AE16} and \citet{Gill17}, while EUV fluxes were taken from \citet{Rib16} and \citet{Bour17} (see also \citet{Bel17}) for Proxima b and the TRAPPIST-1 system respectively.
\end{minipage}
\end{table*}

\begin{figure}
\quad\quad\quad \includegraphics[width=7.18cm]{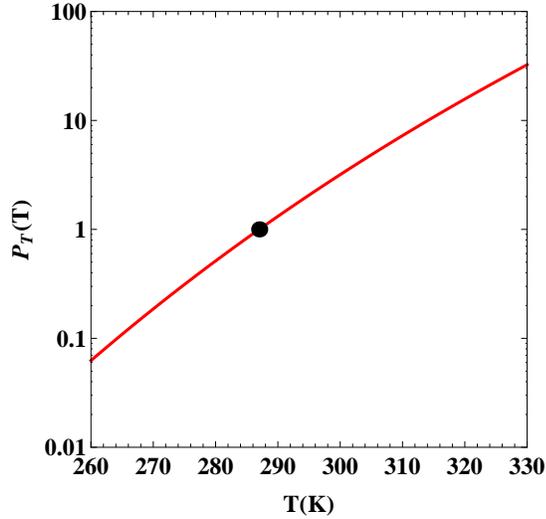} \\
\caption{The likelihood as a function of the planet's average surface temperature $T$. The black dot represents the Earth's value at $T = 287$ K.}
\label{FigT}
\end{figure}

\begin{figure}
\quad\quad\quad \includegraphics[width=7.18cm]{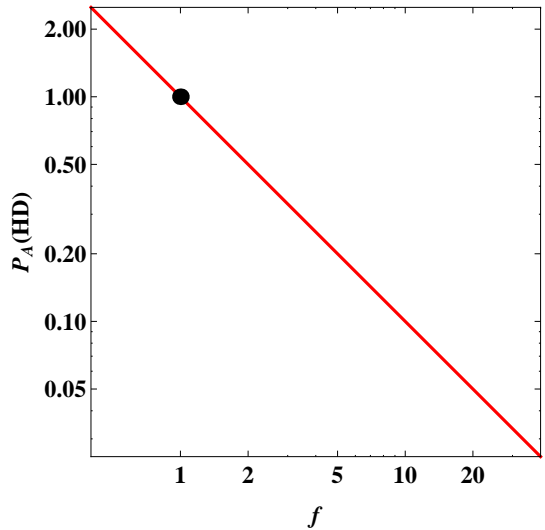} \\
\caption{The likelihood as a function of the normalized EUV flux $f = \langle{F_{EUV}}\rangle/\langle{F_{\oplus}}\rangle$ for an Earth clone. The black dot represents the Earth's position at $f = 1$. } 
\label{FigEUV}
\end{figure}

\begin{figure*}
$$
\begin{array}{cc}
  \includegraphics[width=7.3cm]{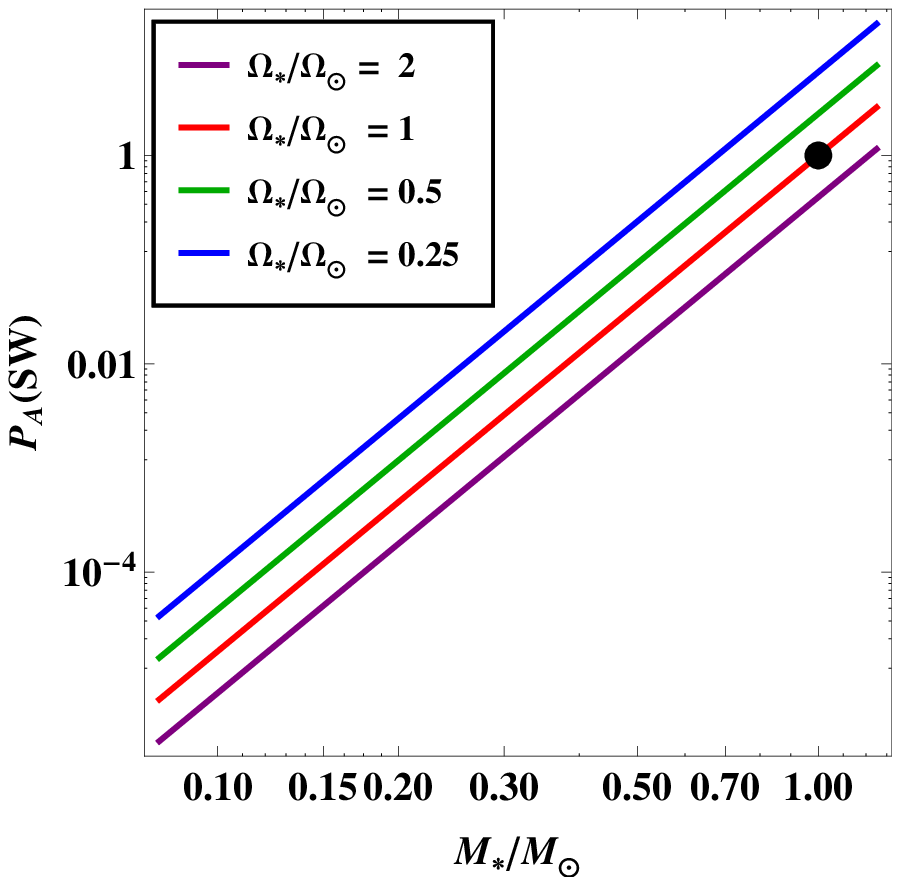} &  \includegraphics[width=7.4cm]{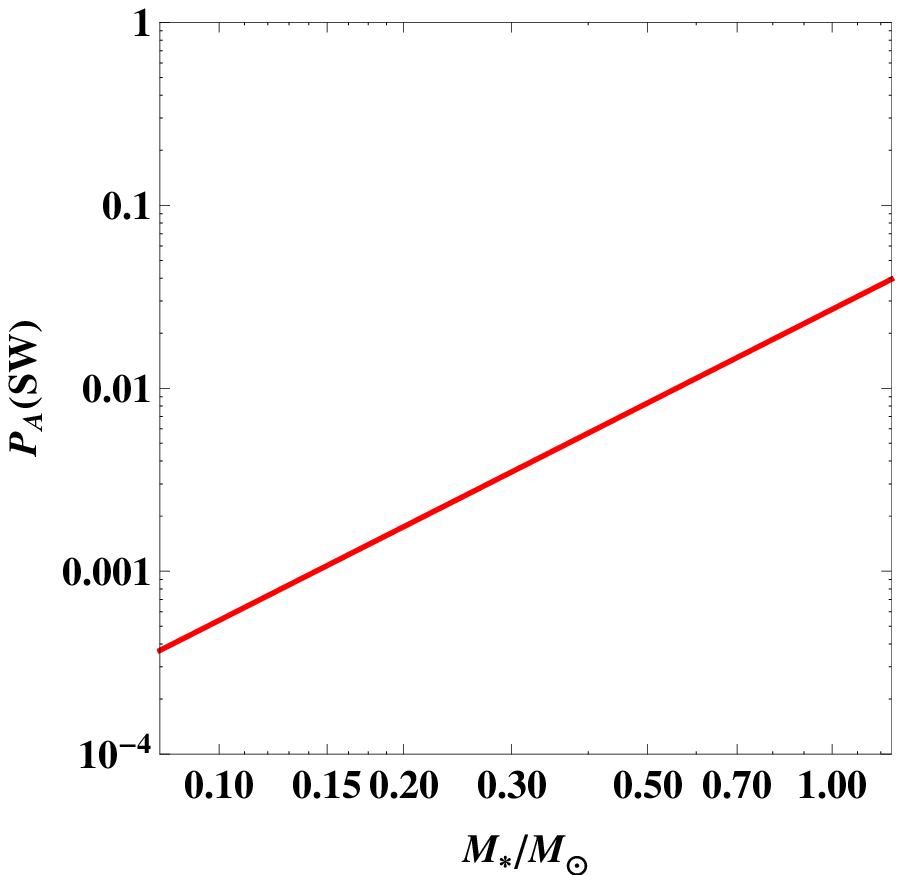}\\
\end{array}
$$
\caption{The likelihood as a function of the stellar mass for an Earth clone. In the left-hand panel, each plot corresponds to a different value of the rotation rate $\Omega_\star$. The black dot signifies the position of the Sun. In the right-hand panel, the likelihood has been plotted for \emph{rapidly rotating} stars, in which case it does not depend on $\Omega_\star$.}
\label{FigM}
\end{figure*}

\section{The likelihood function and its consequences}
We are now in a position to define the overall likelihood function $\mathcal{P}$ from the preceding results,
\begin{equation}
\mathcal{P} = \mathcal{P}_T \cdot \mathcal{P}_A,
\end{equation}
where $\mathcal{P}_T$ is given by Eq. (\ref{TempLike}) and $\mathcal{P}_A$ corresponds to either Eq. (\ref{HDLike}) or Eq. (\ref{SWLike}) depending on the dominant process that drives atmospheric escape.

This function can be used to determine the likelihood of a planet being conducive to life relative to Earth. The chief advantage of our methodology is that nearly all of the parameters are direct observables, or can be deduced indirectly, by means of numerical simulations. The two primary uncertainties involve the surface pressure $P_s$ and the surface temperature $T$. In our subsequent analysis, we shall suppose that the surface pressure is equal to that of the Earth. This still leaves us with $\delta$, which is defined in Eq. (\ref{deldef}). We introduce the ansatz $T = \zeta T_{eq}$, where $T_{eq}$ is the equilibrium temperature and $\zeta$ is a phenomenological parameter that captures the effects of greenhouse warming, snowball dynamics, tidal heating and other feedback mechanisms. In general, $\zeta$ is not constant across all planets, but we adopt this simplifying assumption in order to derive the likelihood function for some of the recently discovered exoplanets in Table \ref{Tab1}. 

We find that the likelihood function $\mathcal{P}_A$ for Proxima b and TRAPPIST-1e is about $1$-$2$ orders of magnitude lower than that of Earth if the atmospheric escape is dominated by hydrodynamic escape. If the atmospheric losses occur due to stellar wind erosion, we conclude that the likelihood is even lower (by three orders of magnitude) for these planets. Another interesting result is that TRAPPIST-1e exhibits a higher likelihood of being habitable (albeit by only a factor of a few) when compared to Proxima b. When dealing with Proxima b, the ostensibly habitable planets of the TRAPPIST-1 system, and other exoplanets in the HZ of M-dwarfs, it is equally important to realize that the timescales for atmospheric loss may be sufficiently short such that $t_\ell < t_{OOL}$, thereby implying that abiogenesis will be non-functional on these planets.

When we consider TRAPPIST-1f and TRAPPIST-1g, the overall likelihood function ends up being zero in our simplified model because $\mathcal{P}_T = 0$ for $T < T_L$. We reiterate that this only represents a coarse-grained, \emph{global} likelihood; in reality, there may exist \emph{locally} favourable temperatures, perhaps near the terminator line, for these two planets enabling $\mathcal{P}_T \neq 0$ to occur in these regions. Our conclusions pertaining to these two planets are also broadly consistent with the results obtained from 3D climate simulations \citep{Wolf17}, although the latter study (as well as our model) does not include the effects of tidal heating that are likely to be significant in the TRAPPIST-1 exoplanetary system.

In Fig. \ref{FigT}, we plot $\mathcal{P}_T$, given by Eq. (\ref{TempLike}), as a function of the planet's surface temperature $T$. This plot can be interpreted as the likelihood of the Earth sustaining a complex biosphere provided that it was characterized by a \emph{steady} surface temperature different from its current value. The presence of the exponential function ensures that the likelihood can vary over two orders of magnitude for a relatively narrow range of $T$. This serves to underscore the fact that a variety of macroecological processes are quite sensitive to the temperature, implying that the latter parameter will clearly play a central role in discussions of planetary habitability. 

Upon inspecting Eq. (\ref{HDLike}) next, we find that it depends on both planetary and stellar parameters. We shall focus on planets that are ``Earth-like'' (but only in a superficial sense), i.e. the surface pressure and radius are chosen to equal the Earth's values. Fig. \ref{FigEUV} depicts the plot of $\mathcal{P}_A(HD)$ as a function of $\langle{F_{EUV}}\rangle$. Physically speaking, this figure quantifies the likelihood function of energy-limited atmospheric escape if the Earth were subjected to varying degrees of EUV flux. 

Finally, we turn our attention to Eq. (\ref{SWLike}). This equation involves a large number of planetary and stellar parameters as well. As before, we consider an ``Earth clone'' with physical parameters equal to that of Earth, namely, $P_s \sim 1$ atm, $R_P \sim R_\oplus$ and $T_{eq} \sim T_{eq,\oplus}$. If we further use $L_\star \propto M_\star^3$ (mass-luminosity relationship), $R_\star \propto M_\star^{0.8}$ \citep{JGBL15} and $a \propto L_\star^{1/2}$ for a fixed value of $T_{eq}$ and the Bond albedo, we end up with the scalings,
\begin{equation} \label{LFFixPPv1}
\mathcal{P}_A(SW) \approx  \left(\frac{\Omega_\star}{\Omega_\odot}\right)^{-1.33} \left(\frac{M_\star}{M_\odot}\right)^{4.76},
\end{equation}
for the likelihood of the Earth's atmosphere to persist - which is a necessary although not sufficient condition for habitability - in the HZ of a different star. Thus, we can immediately see that a slowly-rotating, higher-mass star is more conducive to hosting life on an Earth clone. Note that the above formula is not valid for rapidly rotating stars that yield scaling relations different from the Sun; a similar analysis using the corresponding formula for $\dot{M}_\star$ from \citet{JGBL15} leads to,
\begin{equation} \label{LFFixPPv2}
\mathcal{P}_A(SW) \approx 2.7 \times 10^{-2} \left(\frac{M_\star}{M_\odot}\right)^{1.7},
\end{equation}
which does not exhibit an $\Omega_\star$ dependence. This expression also implies that stars with a higher mass are more likely to host planets in the HZ that are capable of possessing long-term atmospheres.\footnote{The overall lifetime of a star scales with its mass as $M_\star^{-p}$, where $2 < p < 3$ \citep{LBS16}. For a sufficiently massive star, its lifetime will be lower than $t_{OOL}$, implying that the likelihood of life would become zero. Hence, the predicted increase of $\mathcal{P}_A(SW)$ with stellar mass is valid only up to a cutoff value.} We reiterate that Eqs. (\ref{LFFixPPv1}) and (\ref{LFFixPPv2}) are only valid for an Earth clone, and, in reality, $\mathcal{P}_A(SW)$ depends both on stellar and planetary parameters as seen from Eq. (\ref{SWLike}). We have plotted the results from Eqs. (\ref{LFFixPPv1}) and (\ref{LFFixPPv2}) in the two panels of Fig. \ref{FigM}.

\section{Conclusions}
In this paper, we have attempted to address the important question of habitability (metrics) from a more quantitative perspective. We begin with the caveat that our work entails a certain degree of terracentrism with an emphasis on life-as-we-know-it; the associated assumptions cannot be easily bypassed since the Earth is the only planet that is presently known to harbor life. In addition, a genuinely quantitative understanding of habitability is not feasible at this stage given that there exist far too many unknowns \citep{SG16,Task17}.

We have attempted to: (i) draw upon models with strong physical underpinnings, and (ii) present the results in terms of basic physical parameters that can be determined via observations or simulations. We focused on two major physical processes, namely the role of planetary temperature and atmospheric escape. The former was analyzed by means of the MTE, which predicts that many macroecological processes exhibit a Boltzmann factor dependence on the temperature within a certain range. We addressed the possibility of atmospheric escape by considering two different mechanisms, namely hydrodynamic escape and stellar wind induced stripping. 

Our analysis gave rise to a diverse array of conclusions, which have been summarized below:
\begin{itemize}
\item Planets with a warmer mean surface temperature are more conducive to being habitable, insofar macroscopic ecological (as well as evolutionary) processes are concerned, albeit only up to a limited temperature range.
\item Planets with higher surface pressure are, \emph{ceteris paribus}, more likely to be habitable.\footnote{This statement is manifestly valid only in the absence of additional negative consequences arising from a massive atmosphere, such as a runaway greenhouse effect (e.g. Venus), which would otherwise make the planet uninhabitable.} It has also been argued by some authors that abiogenesis on Earth was initiated in a high-pressure environment \citep{MBRR08,PiDa13}, thus indicating that pressure may play a potentially positive role in the origin of life. 
\item For an ``Earth clone'' around a different star, the likelihood of retaining its atmosphere increases with the star's mass, while decreasing with rotation rate and the average ultraviolet (EUV) flux. High doses of ionizing radiation are also likely to have deleterious effects on the functioning of organic molecules and organisms \citep{Dart11}, although the beneficial effects of UV radiation in the context of prebiotic chemistry have been well documented \citep{RaSa16}.
\item On account of the above reasons, Earth-sized planets in the HZ around M-dwarfs are presumably much less likely to be habitable, conceivably by several orders of magnitude, when compared to the Earth-Sun system. Hence, even though M-dwarfs outnumber other stars in our Galaxy, we propose that future searches for life on exo-Earths \citep{HJ10} should prioritize a subset of G and K type stars \citep{KWR93,CuGu16}. This may also explain why we live on a terrestrial planet currently in the HZ of the Sun, and not one that is in the HZ of an M-dwarf in the cosmic future \citep{LBS16,HKW17}.
\item More specifically, Proxima b and TRAPPIST-1e yield a much lower likelihood of being habitable (by $2$-$4$ orders of magnitude), with respect to atmospheric escape mechanisms, when compared to Earth. Furthermore, TRAPPIST-1e appears to be more conducive to hosting life, while only by a factor of a few, with respect to Proxima b if one makes the further assumption that their host stars have a similar stellar mass-loss rate.
\item If the assumptions in this paper are valid, it seems plausible that TRAPPIST-1f and TRAPPIST-1g are incapable of sustaining life across a significant fraction of the surface, although the possibility of local life-bearing zones and life seeded by means of panspermia \citep{Ling16,LL17} cannot and ought not be ruled out.
\item As most of the stellar and planetary parameters are time-dependent, the likelihood functions are also implicitly dynamical.\footnote{For instance, the stellar mass-loss rate in Eq. (\ref{tSW}) changes significantly, by several orders of magnitude, over time for solar-like stars \citep{WMZL}.} This fact is fully consistent with the notion that planetary habitability (and sustainability) evolves over time \citep{FraSu14}.
\item Many of these findings, after suitable reinterpretation, are also applicable to habitable exomoons which may outnumber habitable exoplanets \citep{Hell14}.
\end{itemize}
We end by cautioning that our proposed methodology is by no means complete, since important (first-order) feedback mechanisms - such as the complex, nonlinear and adaptive interplay between life and planetary habitability \citep{Lev68,LoMa74,Lev98,AFS04,CL16} - have \emph{not} been included herein \citep{LC12,Ehl16,Jud17}. We also wish to point out that we have altogether neglected second-order effects in this study. For instance, it has been suggested that a larger surface area facilitates higher biodiversity \citep{Ros95}, thereby making the planet superhabitable \citep{HeArm14}. However, it is important to recognize that the radius (and hence the area) of super-Earths is quite constrained \citep{Rog15,CK17}, thereby implying that this effect qualifies as an $\mathcal{O}(1)$ contribution.

As our work is founded on physical and mechanistic considerations inclined towards universality, we advocate the adoption of such an approach in future studies that seek to quantify the likelihood of exoplanets hosting complex and long-lived biospheres. Pursuing such lines of enquiry may also prove to be a natural and timely means of investigating the dependence of abiogenesis, at least for life-as-we-know-it, on biochemical (possibly even planetary and stellar) parameters by means of associated paradigms \citep{GoWo11,PPS13,DaWa16}, thereby complementing previous probabilistic estimates \citep{LD02,Cart08,ST12,SC16}. Studies along these lines would lead us towards a resolution of the fundamental question as to whether life (and intelligence) in the universe is an extremely rare phenomenon \citep{Gay64,Mon71,WaB00,Mor03} or an inevitable `cosmic imperative' \citep{Duve95,Duve11,RNB13}.\footnote{With regards to the latter viewpoint, the reader may also consult the likes of \citet{Kau95,MS07,BVBF07,Flo14,BaSM16} that explore similar perspectives.}

\acknowledgments
We thank Chuanfei Dong, John Forbes, Jeffrey Linsky, Sukrit Ranjan, John Raymond, Ed Turner, and the referee for their helpful comments concerning the paper. This work was supported in part by a grant from the Breakthrough Prize Foundation for the Starshot Initiative.


\end{document}